\documentclass[10pt,conference]{IEEEtran}

\usepackage{url}
\usepackage{array}
\usepackage{xcolor}
\usepackage{cancel}
\usepackage{siunitx}
\usepackage{multirow}
\usepackage{graphicx}
\usepackage{textcomp}
\usepackage{multirow}
\usepackage{stfloats}
\usepackage{algorithm}
\usepackage[T1]{fontenc}
\usepackage[noadjust]{cite}
\usepackage[10pt]{moresize}
\usepackage[utf8]{inputenc}
\usepackage{mathtools, cuted}
\usepackage{amsmath,amsfonts}
\usepackage[noend]{algpseudocode}
\usepackage{soulutf8}

\IEEEoverridecommandlockouts

\graphicspath{{./figs/}}

\newlength{\textfloatsepsave} 
\begin{document}

\title{Resynthesis-based Attacks Against Logic Locking}
    
%\author{Felipe Almeida, Levent Aksoy,~\IEEEmembership{Member,~IEEE,} Quang-Linh~Nguyen,~\IEEEmembership{Student Member,~IEEE,}, Sophie~Dupuis,~\IEEEmembership{Member,~IEEE,}, Marie-Lise~Flottes,~\IEEEmembership{Member,~IEEE,} and~Samuel~Pagliarini,~\IEEEmembership{Member,~IEEE}

    \author{
	\IEEEauthorblockN{Felipe~Almeida\IEEEauthorrefmark{2},
			Levent~Aksoy\IEEEauthorrefmark{2},
			Quang-Linh~Nguyen\IEEEauthorrefmark{3}, 
			Sophie~Dupuis\IEEEauthorrefmark{3},
			Marie-Lise~Flottes\IEEEauthorrefmark{3} and	Samuel Pagliarini\IEEEauthorrefmark{2}}
		\IEEEauthorblockA{\IEEEauthorrefmark{2}Department of Computer Systems, Tallinn University of Technology, Tallinn, Estonia\\
			Email: \{felipe.almeida, levent.aksoy,  samuel.pagliarini\}@taltech.ee}
		\IEEEauthorblockA{\IEEEauthorrefmark{3}LIRMM, University of Montpellier, Montpellier, France\\
			Email: \{quang-linh.nguyen, sophie.dupuis, marie-lise.flottes\}@lirmm.fr}
   \vspace*{-1cm}

    \thanks{This work has been partially conducted in the project ``ICT programme'' which was supported by the European Union through the European Social Fund. It was also partially supported by European Union's Horizon 2020 research and innovation programme under grant agreement No 952252 (SAFEST).}}

    \maketitle

    %\vspace*{-4cm}

%\thanks{L.~Aksoy, F.~Almeida, J.~Raik, and S.~Pagliarini are with the Department of Computer Systems, Tallinn University of Technology, Tallinn, Estonia (e-mail: \{levent.aksoy, felipe.almeida, jaan.raik, samuel.pagliarini\}@taltech.ee.)}

%\thanks{Q.-L.~Nguyen, M.-L.~Flottes, and S.~Dupuis are with LIRMM, University of Montpellier, CNRS, Montpellier, France (e-mail: \{quang-linh.nguyen, marie-lise.flottes, sophie.dupuis\}@lirmm.fr.)}

    \begin{abstract}
        Logic locking has been a promising solution to many hardware security threats, such as intellectual property infringement and overproduction. Due to the increased attention that threats have received, many efficient specialized attacks against logic locking have been introduced over the years. However, the ability of an adversary to manipulate a locked netlist prior to mounting an attack has not been investigated thoroughly. This paper introduces a resynthesis-based strategy that utilizes the strength of a commercial electronic design automation (EDA) tool to reveal the vulnerabilities of a locked circuit. To do so, in a pre-attack step, a locked netlist is resynthesized using different synthesis parameters in a systematic way, leading to a large number of functionally equivalent but structurally different locked circuits. Then, under the oracle-less threat model, where it is assumed that the adversary only possesses the locked circuit, not the original circuit to query, a prominent attack is applied to these generated netlists collectively, from which a large number of key bits are deciphered. Nevertheless, this paper also describes how the proposed oracle-less attack can be integrated with an oracle-guided attack. The feasibility of the proposed approach is demonstrated for several benchmarks, including remarkable results for breaking a recently proposed provably secure logic locking method and deciphering values of a large number of key bits of the CSAW'19 circuits with very high accuracy. 
    \end{abstract}
    
    \begin{IEEEkeywords}
		Logic locking, resynthesis, EDA tools, oracle-less and oracle-guided attacks.
	\end{IEEEkeywords}

	\section{Introduction}

Due to the globalized integrated circuit (IC) supply chain, serious security threats, such as hardware Trojans, piracy, overbuilding, reverse engineering, and counterfeiting, have emerged~\cite{rostami14}. Many defense techniques, such as watermarking~\cite{kahng98}, digital rights management~\cite{alkabani07}, metering~\cite{koushanfar07}, and logic locking~\cite{Dupuis2019}, have been introduced over the years to deal with these threats. Among those, logic locking stands out by being a well-established technique and by offering protection against a diverse array of adversaries~\cite{yasin17}. Logic locking inserts additional logic driven by key bits so that the circuit behaves as expected only when the secret key is applied. 

On the other hand, many efficient attacks have been introduced to overcome the defenses built by logic locking~\cite{Azar2019}. However, the impact of an electronic design automation (EDA) tool on the manipulation of the locked netlist \textbf{before} performing an attack has not been investigated thoroughly. In this work, we explore if EDA tools can be used to make a locked circuit vulnerable to existing logic locking attacks. Thus, the main contributions of this work are three-fold: (i)~we introduce a resynthesis procedure that is a \textbf{pre-attack} step, where functionally equivalent but structurally different locked circuits are generated by resynthesizing the original locked circuit using different optimization parameters and delay constraints in order to create structural vulnerabilities that can be exploited by existing attacks; (ii)~we present an oracle-less (OL) \textbf{resynthesis-based attack}, which applies the prominent SCOPE attack~\cite{alaql2021} to these resynthesized circuits and gathers all its solutions to discover the secret key; (iii)~we show that our OL attack can be \textbf{combined} with a traditional oracle-guided (OG) attack for further improving the number of correctly deciphered key bits. The last contribution is essential, since we consider circuits from the CSAW'19 contest -- these circuits compound the use of two logic locking techniques at the same time. 

The main finding of this work is that the use of many resynthesized locked circuits enables us to discover values of more key bits, and even the whole key, when compared to a single attack mounted on the original locked netlist. 

%Moreover, we show that the resynthesis-based attack can discover the secret key of a design locked by a recently proposed provably secure logic locking (PSLL) technique~\cite{nguyen2021}.

The remainder of this paper is organized as follows: Section~\ref{sec:background} presents the background concepts and related work. The resynthesis process and the proposed attacks are described in Section~\ref{sec:methodology}. Experimental results are given in Section~\ref{sec:results}. Finally, Section~\ref{sec:conclusions} concludes the paper.

	\section{Background}
\label{sec:background}

\subsection{Logic Locking and Threat Models}

\begin{figure*}[t]
	\centerline{\includegraphics[width=18.0cm]{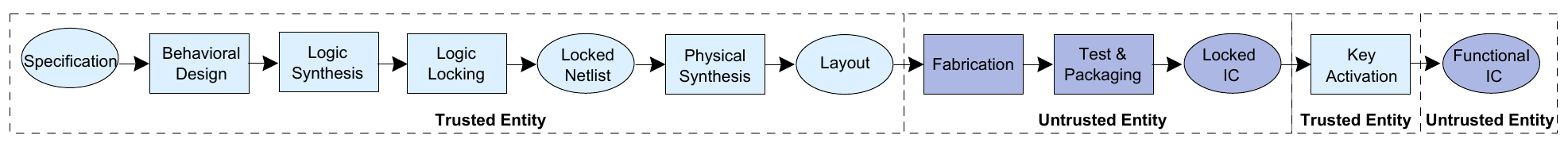}}
        \vspace*{-4mm}
	\caption{Conventional logic locking in the IC design flow (adapted from~\cite{yasin17}).}
	\label{fig:icflow}
\end{figure*}

The procedure of logic locking is applied at the gate level in the IC design flow, as shown in Fig.~\ref{fig:icflow}. Note that the layout of the locked circuit is sent to the foundry without revealing the secret key. After the locked IC is produced and delivered to the design house, the values of the secret key are stored in a tamper-proof memory, before the functional IC is sent to the market. 

It is assumed that the gate-level netlist of the locked circuit can be obtained directly by an untrusted foundry or by \mbox{reverse-engineering} a functional IC obtained from the open market. An adversary can also use the functional IC programmed with the secret key as an oracle to apply inputs and observe outputs. Thus, in logic locking, there are generally two threat models: OL and OG. In the OL threat model, only the gate-level netlist of the locked circuit is available to the adversary. The adversary has both the netlist of the locked circuit and the functional IC in the OG threat model. 

%The attack proposed in this paper assumes the more restrictive OL threat model.

\begin{figure}[t]
	\centerline{\includegraphics[width=9.0cm]{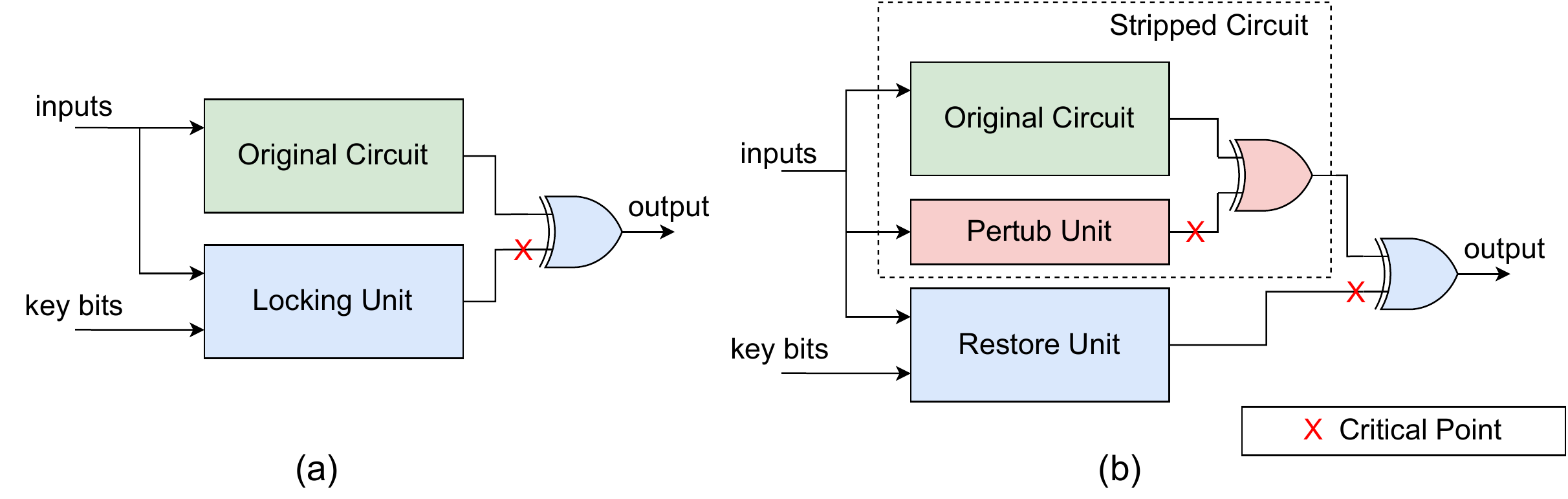}}
        \vspace{-4mm}
	\caption{SAT-resilient logic locking methods: (a)~SFLT; (b)~DFLT.}
	\label{fig:classification}
\end{figure}

\vspace{-4mm}
\subsection{Related Work}

After the introduction of random logic locking (RLL) using {\sc xor/xnor} gates in~\cite{roy2008}, earlier work focused on different types of key gates, such as {\sc and/or}, multiplexors, and \mbox{look-up} tables, taking into account the hardware complexity of the locked circuit~\cite{Dupuis2019}. However, the OG satisfiability (SAT)-based attack~\cite{subramanyan2015} overcame all the defenses existing at that time. Note that the SAT-based attack iteratively finds differentiating input patterns (DIPs) that rule out wrong keys. To thwart the \mbox{SAT-based} attack and its variants, circuits are locked using a point function that sets a limit on the number of wrong keys which a DIP can eliminate, forcing these attacks to explore an exponential number of queries~\cite{yasin17, xie19, shakya19, sengupta20, nguyen2021}. 

The SAT-resilient methods can be categorized into two groups: single-flip locking technique (SFLT) and double-flip locking technique (DFLT), as shown in Fig.~\ref{fig:classification}. An SFLT has only one critical point, which is responsible to corrupt a protected output under a specific input pattern. In this category, SARLock~\cite{yasin2016} adds a comparator and a masking circuit connected with the original netlist in a way that it generates a corruption on one input pattern. Anti-SAT~\cite{xie19} utilizes two complementary {\sc and} gate trees, whose output is merged with the original circuit. CASLock~\cite{shakya19} is based on the same concept of Anti-SAT, however it uses both {\sc and} and {\sc or} gates. SKG-Lock~\cite{nguyen2021} uses decoy key bits and provides a tunable output corruption. Note that SFLTs are susceptible to removal attacks~\cite{xu17, yasin20, sengupta21}. If an attacker can identify this single critical point, he/she can split the design into a recovered netlist (original) and the locking unit. 

A DFLT has two critical points, one that connects the original netlist with a perturbation unit and another one that connects the output of the stripped circuit with the restore unit. Under this category, stripped functionality logic locking (SFLL)~\cite{yasin17, sengupta20} initially corrupts an output based on an input combination in the perturbation unit and then, corrects this output only when the secret key is applied in the restore unit. Note that a removal attack becomes inefficient for a DFLT, since the original circuit is mixed with the perturbation unit, even though it can easily identify the restore unit. However, there exist efficient structural attacks developed for DFLTs~\cite{sirone19, yang19, zhaokun2021, limaye22}. 

%A recently proposed attack, called Valkyrie~\cite{limaye22}, can identify critical points in the design and recover the original netlist removing the locking unit for SFLT or the perturb and restore units for DFTL. 

Alternative locking techniques have also been introduced. In~\cite{liu20}, a technique, which has more than two critical points, called the multi-flip locking technique (MFLT), was proposed. However, it leads to a significant increase in area, power dissipation, and delay when compared to other techniques. Compound logic locking techniques were proposed to overcome the main drawback of a SAT-resilient technique, i.e., its low output corruptibility as can be observed in Fig.~\ref{fig:classification}, by locking a design using both low and high output corruptibility techniques, such as SFLL and RLL, respectively~\cite{tan20}. Recently, efficient attacks have also been introduced against compound logic locking~\cite{john20, nimisha21}. 

Moreover, the OL attacks explore patterns in the structure of a locked netlist using statistical analysis~\cite{li19,zhang19,alaql2021}. For example, the SCOPE attack~\cite{alaql2021} is an unsupervised constant propagation technique, which analyzes each key bit of the locked design for critical features that can reveal its correct value after it is assigned to logic 0 and 1 value. These critical features include area, power dissipation, delay, and many other circuit characteristics obtained by a synthesis tool. These features are analyzed using linear regression and machine learning based clustering.
	\section{Proposed Resynthesis-based Attack}
\label{sec:methodology}

This section describes our resynthesis-based attack in detail. We initially introduced the pre-attack stage, where the locked circuit is resynthesized using different synthesis parameters, leading to a large number of structurally different netlists with the same functionality \cite{tool2023}. Then, we present the OL attack that utilizes these resynthesized netlists in order to find the secret key. Finally, in order to handle the compound logic locking efficiently, we present its modified version, where our proposed OL attack cooperates with an OG attack.

\subsection{The Pre-attack Step: Resynthesis of the Locked Netlist}
\label{subsec:resynth}

The locked circuit is synthesized multiple times using a different script each time, where the synthesis parameters are explored in a systematic way. We use the following parameters to increase the number of resynthesized locked circuits:

\textit{Synthesis Effort}: In a synthesis tool, logic optimizations can be applied with different efforts at different synthesis stages. This flexibility enables a designer to explore the trade-off between the quality of results and run time. The following efforts are considered at the given synthesis stage: low, medium, and high at generic transformations (\textit{syn\_gen}); low, medium, and high at mapping (\textit{syn\_map}); and low, medium, high, and extreme at optimization (\textit{syn\_opt}).

\textit{Delay Constraint}: To meet performance targets, delay constraints are used to guide the synthesis tool. We initially resynthesize the locked circuit without a delay constraint and find the delay of its critical path, i.e., $dcp$. Then, in an interval between 0 and $dcp$, $d-1$ points, which are computed as $(dcp/d)i$ with $1 \leq i \leq d-1$, are set as delay constraints. Note that $d$ is set to 5 in order to generate a large number of resynthesized circuits. Even though some delay constraints are impossible to meet, the synthesis tool \emph{always} generates a netlist equivalent to the original one in terms of functionality.

\textit{Maximum Transition}: The transition time of a net in a circuit is defined as the longest time required for its driving pin to change its logic value. We choose the maximum transition value to be 5\%, 10\%, and 15\% of the delay constraint for all the nets in the locked circuit to explore different resynthesized circuits.

\textit{Key Constraints}: To direct the synthesis tool to work intensively on the paths that contain the keyed logic, a delay constraint, which is impossible to be satisfied, can also be used. In this case, we force the delay between all key bits and all primary outputs to be 1\,ps.

Thus, the combination of parameters given above generates $3 \times 3 \times 4 \times 5 \times 3 \times 2 = 1080$ netlists. We eliminate the resynthesized circuits with identical characteristics and keep only the unique ones. Additionaly, we prevent the use of {\sc xor/xnor} gates, which can be problematic for the SCOPE attack, during technology mapping. Note that our resynthesis methodology aims to generate different versions of the locked circuit, making it more vulnerable to existing attacks. Thus, any existing attack, either OL or OG, may potentially benefit from this pre-attack strategy to discover the secret key.

\subsection{Attacks on the Resynthesized Netlists}

Time-efficient attacks are chosen in order to handle a large number of resynthesized circuits. In our OL resynthesis-based attack, SCOPE~\cite{alaql2021} is used to predict the values of key bits. In its modified version developed for compound logic locking, a query attack is used to find the values of key bits in a deterministic way.

\subsubsection{Proposed OL Attack}

SCOPE is applied to each resynthesized locked circuit and a solution is found. Note that this solution may return a logic 0, 1, or an unknown value for a key bit. Then, the values of key bits deciphered for each netlist are merged into a single solution that represents the overall guess. To do so, for each key bit, $k_i$ with $1 \leq i \leq p$, where $p$ denotes the number of key bits, we initially count the number of solutions, where $k_i$ is deciphered as logic 0 and 1, denoted as $dk_{i}^{0}$ and $dk_{i}^{1}$, respectively. Then, if $dk_{i}^{0} > dk_{i}^{1}$ or $dk_{i}^{1} > dk_{i}^{0}$, the value of $k_i$ is determined to be 0 or 1, respectively. Otherwise, in the case of a tie, the value of $k_i$ is decided to be unknown. 

\subsubsection{Proposed OG Attack}

In order to handle a large number of resynthesized netlists efficiently, we introduce a \mbox{SAT-based} query attack, which can determine the actual values of individual key bits. Note that traditional SAT-based attacks rather attempt to find the whole secret key, which increases the computational effort significantly. In this attack, we initially find queries, i.e., values of inputs of the oracle circuit, using two techniques. The first technique uses the ATPG tool Atalanta~\cite{atalanta} to find test patterns for the stuck-at-fault of each key bit on the locked circuit and stores the values of the related primary inputs as queries. The aim is to find input patterns that can propagate each key bit to a primary output, making it observable. The second technique finds queries randomly. The aim is to find input patterns that may make multiple key bits observable. In our experiments, we generate a total of $2p$ queries, where $p$ denotes the number of key bits. 

Then, we describe the locked circuit in a conjunctive normal form (CNF) formula $\mathbb{C}$ by expressing each gate in its CNF. Each query is applied to the oracle and the values of primary outputs are obtained. Then, the related input and output values are assigned to the associated nets in the locked circuit, the constant values of these nets are propagated, and the Boolean equations including key bits are derived in a CNF formula $\mathbb{E}$. The SAT problem including the locked circuit in CNF, i.e., $\mathbb{C}$, is augmented with these equations, i.e., $\mathbb{C} = \mathbb{C} \land \mathbb{E}$. After all the queries are considered, the SAT problem $\mathbb{C}$ is solved using a SAT solver and the values of key bits are determined. Note that the locked circuit with the found values of key bits behaves exactly the same as the oracle under the given queries, but not under all possible input values. Hence, these key values are not guaranteed to be the values of the secret key. 

However, the value found for a key bit can be proved if it is indeed equal to the actual value of the related bit in the secret key using the concept of \textit{proof by contradiction}. To do so, for each key bit, the complement of its found value is added into $\mathbb{C}$ and the SAT solver is run. If there exists no solution to $\mathbb{C}$, i.e., the SAT problem is unsatisfiable, the value of the related key bit is proven to be the one in the found solution. 

\begin{figure}[t]
    \centering
    \includegraphics[width=8.8cm]{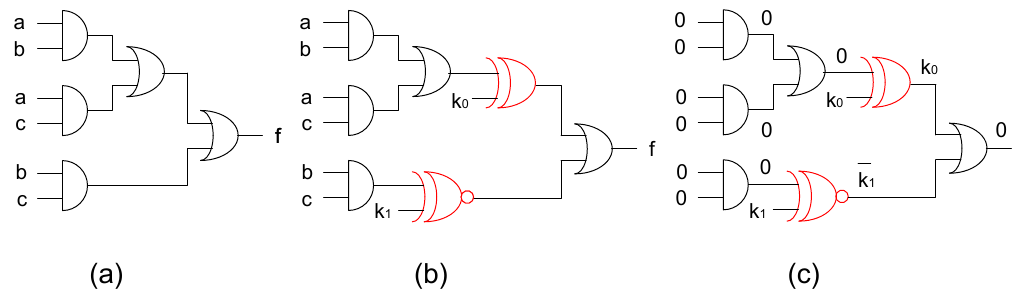}
    \vspace{-4mm}
    \caption{(a)~Majority circuit; (b)~Locked majority circuit; (c)~Constant propagation on the locked majority circuit.}
    \label{fig:majority}
    \vspace{-4mm}
\end{figure}

As a simple example, consider the majority circuit in Fig.~\ref{fig:majority}(a) and suppose that it is locked using {\sc xor/xnor} gates as given in Fig.~\ref{fig:majority}(b). Assume that a query is found as $abc=000$ and thus, the value of its output $f$ is obtained as 0 using the oracle. After propagating these values on the locked circuit as shown in Fig.~\ref{fig:majority}(c), a Boolean equation $k_0 \lor \overline{k_1} = 0$, i.e., $\overline{k_0} \land k_1$ in CNF, is obtained. In the SAT solution, the key bit values are found as $k_0k_1=01$. Note that these are the proven key values since a SAT solver guarantees that there exists no solution to the SAT problem $\mathbb{C}$, which is extended by either $k_0=1$, i.e., $k_0$ in CNF, or $k_1=0$, i.e., $\overline{k_1}$ in CNF, due to a conflict with the found Boolean equation, i.e., $\overline{k_0} \land k_1$ in CNF.

The query attack is run on all the resynthesized circuits and the proven values of key bits in each netlist are combined into a single solution. Note that the query attack is developed in Perl and is equipped with the incremental SAT solver CaDiCaL~\cite{biere20}.

Finally, the solution of the OG resynthesis-based attack is determined after merging the solution of the SCOPE attack over all resynthesized circuits into that of the query attack on all resynthesized circuits without changing the proven values of key bits. 

	\section{Experimental Results}
\label{sec:results}

This section initially presents the results of the proposed OL resynthesis-based attack on the ISCAS'85 circuits~\cite{iscas1985} and then, those of the OG resynthesis-based attack on the CSAW'19 circuits~\cite{tan20} including compound logic locking.

\begin{table}[t]
  \centering
  \footnotesize
  \caption{Details of the ISCAS'85 Circuits.}
  \begin{tabular}{|@{\hskip3pt}l@{\hskip3pt}|c@{\hskip3pt}c@{\hskip3pt}c|c@{\hskip3pt}|c@{\hskip3pt}c@{\hskip3pt}c@{\hskip3pt}c@{\hskip3pt}|}
    \hline
    \multirow{3}{*}{Circuit} & \multicolumn{3}{c|}{\multirow{2}{*}{Original Netlist}} & \multicolumn{5}{c|}{Locked Netlist} \\ 
    \cline{5-9}
    & & & & \multirow{2}{*}{$p$} & Anti-SAT & CASLock & SFLL & SKG-Lock\\ 
    \cline{2-4} \cline{6-9}
    & \#in & \#out & \#gates & & \#gates & \#gates & \#gates & \#gates\\
    \hline \hline
    c2670 & 157 & 64  & 1193 & 64 & 1321 & 1320& 1421 & 1401\\
    c3540 & 50  & 22  & 1669 & 32 & 1733 & 1732& 1783 & 1773\\
    c5315 & 178 & 123 & 2307 & 64 & 2435 & 2434& 2523 & 2514\\
    c6288 & 32  & 32  & 2416 & 32 & 2480 & 2479& 2531 & 2516\\
    c7552 & 206 & 105 & 3512 & 64 & 3640 & 3639& 3729 & 3713\\
    \hline
  \end{tabular}
  \label{tab:iscas85}
  \vspace{-4mm}
\end{table}

\begin{table*}[t]
  \centering
  \footnotesize
  \caption{Results of Resynthesized Locked ISCAS'85 Circuits.}
  \begin{tabular}{|c|l|c|c|c|c|c|}
  \hline
    {Technique} & \multicolumn{1}{c|}{Details} & {c2670} & {c3540} & {c5315} & {c6288} & {c7552}\\
    \hline \hline
    \multirow{5}{*} {Anti-SAT} & \#unique & 480 & 537 & 464 & 498 & 439\\
    & area & 2357 & 2803 & 4112 & 7265 & 5387\\
    & delay& 504 & 818 & 663 & 2144 & 694\\
    & power& 5518 & 4934 & 4297 & 9403 & 7479\\
    & time & 17h14m51s & 1d05h56m12s & 1d09h56m22s & 3d20h50m46s & 1d16h01m13s\\
    \hline 
    \multirow{5}{*} {CASLock} & \#unique & 473 & 449 & 488 & 410 & 479\\
    & area & 2359 & 3112 & 4173 & 7739 & 5337\\
    & delay & 513 & 874 & 650 & 2146 & 676\\
    & power & 5170 & 3304 & 3852 & 10693 & 6765\\
    & time & 15h29m56s & 1d11h02m52s & 1d06h52m54s & 4d03h12m29s & 1d16h06m52s\\
    \hline
    \multirow{5}{*} {SFLL} & \#unique & 468 & 484 & 477 & 523 & 504 \\
    & area & 2817 & 3444 & 4326 & 7646 & 5340 \\
    & delay & 481 & 870 & 697 & 2144 & 604 \\
    & power & 6189 & 6337 & 9053 & 12115 & 11320 \\
    & time & 13h13m23s & 1d47m51s & 21h57m07s & 2d22h15m07s & 22h40m29s \\
    \hline
    \multirow{5}{*} {SKG-Lock} & \#unique & 521 & 541 & 507 & 527 & 521 \\
    & area & 2673 & 2773 & 4646 & 6293 & 4774 \\
    & delay & 936 & 986 & 782 & 2093 & 874 \\
    & power & 3881 & 3831 & 8160 & 7201 & 7822 \\
    & time & 22h22m01s & 1d08h8m27s & 1d03h56m15s & 2d14h29m32s & 1d04h19m \\
    \hline
  \end{tabular}
  \label{tab:resynth-iscas85}
  \vspace*{-4mm}
\end{table*}

\subsection{Results on the ISCAS'85 Circuits}
\label{subsec:iscas85}

As the first experiment set, five ISCAS'85 circuits were considered. Table~\ref{tab:iscas85} presents their details. For our experiments, these circuits were locked by the Anti-SAT~\cite{xie19}, CASlock~\cite{shakya19}, SFLL~\cite{yasin17}, and SKG-Lock~\cite{nguyen2021} techniques. Note that while Anti-SAT and SFLL were taken from the NEOS tool~\cite{neos}, SKG-Lock was provided by its developer, and CASLock was implemented by ourselves. Table~\ref{tab:iscas85} also presents details of the locked circuits. Note that the number of keys, i.e., $p$, was determined based on the number of inputs and overhead of the locking technique, and circuit characteristics, i.e., the number of inputs, outputs, and gates, were taken from the gate-level netlist.

Observe from Table~\ref{tab:iscas85} that all logic locking techniques lead to circuits with a number of gates close to each other, whereas the one locked by SFLL has a slightly large number of gates. Besides, the overhead on the number of gates in circuits locked by SFLL varies from 4.7\% to 19.1\% when compared to original circuits.

In the following subsections, we present the results of the resynthesis process and OL resynthesis-based attack, analyze the impact of synthesis parameters on the performance of the resynthesis process and SCOPE attack, and introduce improvements to the run-time of the resynthesis process.

\subsubsection{Resynthesis of the Locked ISCAS'85 Circuits}

The resynthesis is performed by Cadence Genus with a commercial 65\,nm standard cell library, and the whole process is automated in a Perl script. Table~\ref{tab:resynth-iscas85} presents the resynthesis results of locked circuits. In this table, \textit{unique} denotes the number of unique locked netlists out of 1080 generated netlists and \textit{area}, \textit{delay}, and \textit{power} stand respectively for the average values of total area in $\mu m^2$, delay in the critical path in $ps$, and total power dissipation in $\mu W$ on the unique locked netlists. Finally, \textit{time} is the total run time of the resynthesis process. The resynthesized netlists were generated on a computing server with Intel Xeon processing units at 3.9\,GHz and a total of 1\,TB memory.

\begin{figure*}[t]
	\centering
	\parbox{5.8cm}{\centerline{\includegraphics[width=6.5cm]{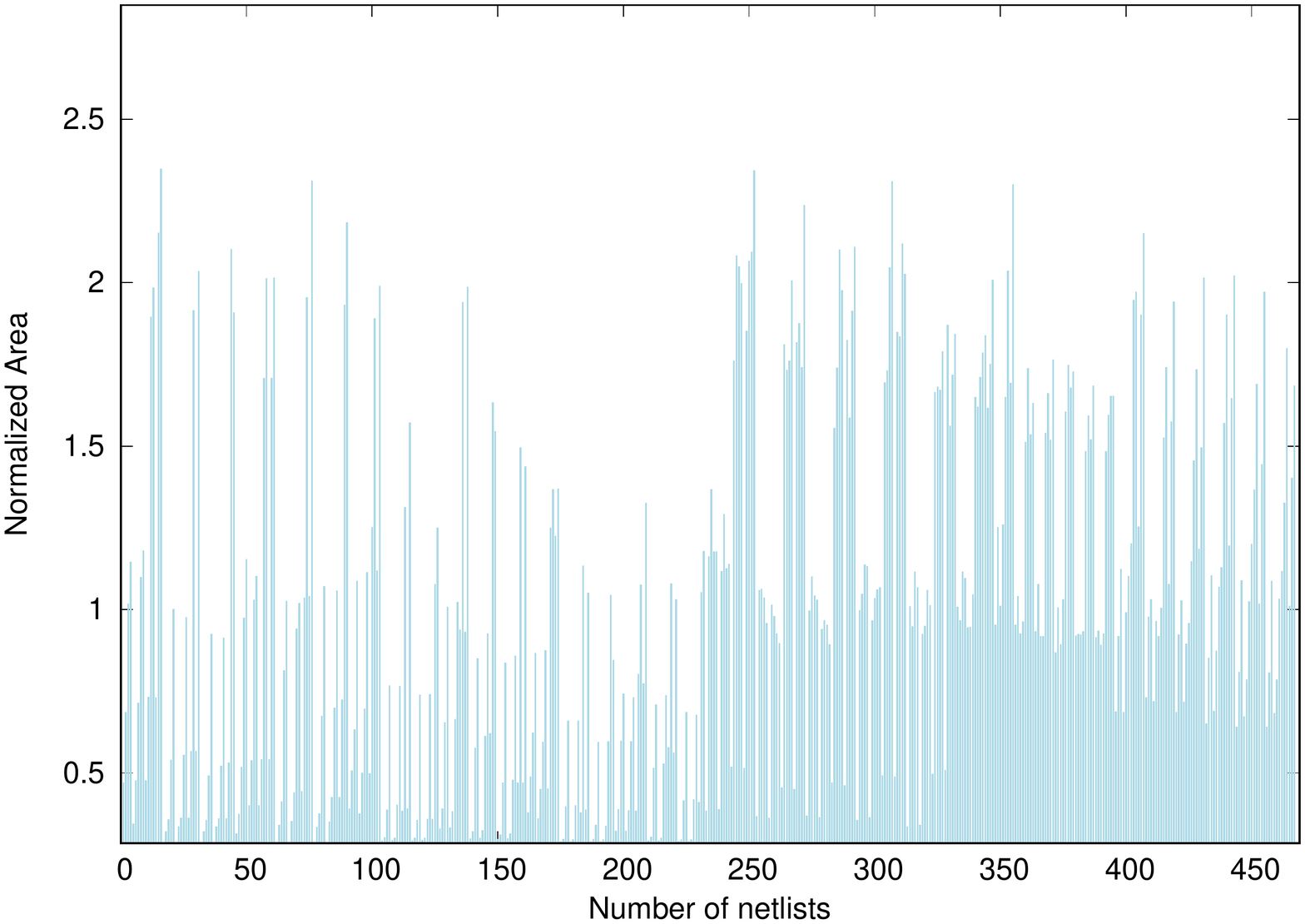}}}\
	\parbox{5.8cm}{\centerline{\includegraphics[width=6.5cm]{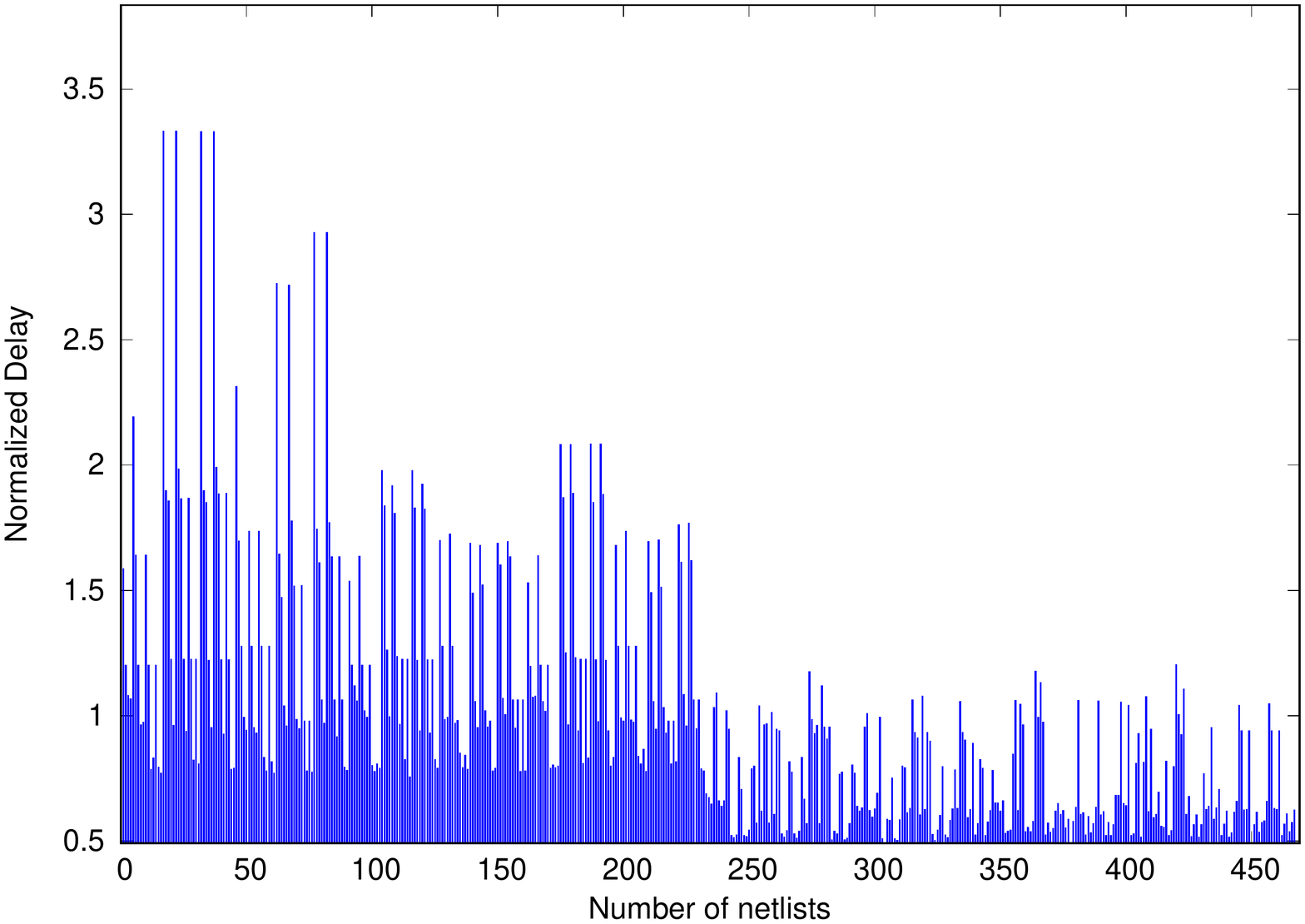}}}\
	\parbox{5.8cm}{\centerline{\includegraphics[width=6.5cm]{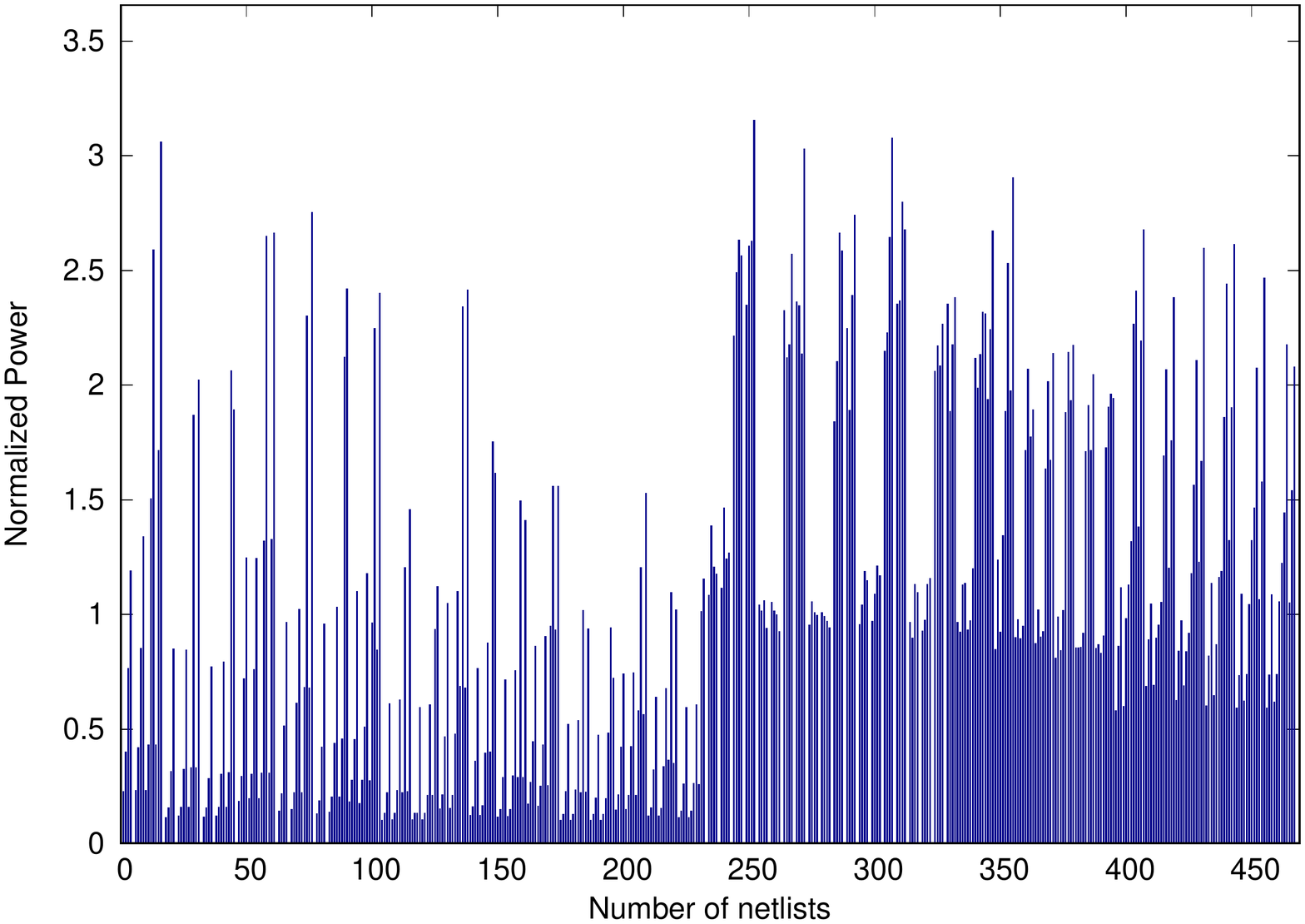}}}\

	\parbox{5.8cm}{\centerline{\footnotesize (a)}}\
	\parbox{5.8cm}{\centerline{\footnotesize (b)}}\
	\parbox{5.8cm}{\centerline{\footnotesize (c)}}\
	\vspace*{-3mm}
	\caption{Normalized complexity of resynthesized netlists of the c2670 circuit locked by SFLL: (a)~area; (b)~delay; (c)~power.}  
	\label{fig:resynth-iscas85}
        \vspace*{-4mm}
\end{figure*}

Observe from Table~\ref{tab:resynth-iscas85} that the number of unique netlists is less than half of the total number of generated netlists, i.e., 540, except the \textit{c3540} circuit locked by SKG-Lock. Note that Anti-SAT, CASLock, and SFLL lead to fewer unique netlists when compared to SKG-Lock, which is mainly because the logic added by these techniques is more compact than that added by SKG-Lock, which uses a chain of {\sc and} gates. We note that the synthesis tool consumes a large amount of time to fulfill a delay constraint that is impossible to meet, such as strict delay constraints and key constraints described in Section~\ref{subsec:resynth}. Hence, the run-time of the resynthesis process depends on the locked circuit and the logic locking technique, and more importantly, if there exists enough room for the synthesis tool to satisfy the constraints.

In order to illustrate the diversity of resynthesized netlists, the \textit{c2670} circuit locked by SFLL is considered. Fig.~\ref{fig:resynth-iscas85} presents the area, delay, and power dissipation of each unique netlist, normalized by their average values given in Table~\ref{tab:resynth-iscas85}. Observe that resynthesis generates circuits significantly different from each other in terms of hardware complexity. The standard deviation on area, delay, and power dissipation values of all these netlists are computed as 1578, 235, and 4964, respectively. Note also that in this figure, the netlists after instance number 232 have a distinct profile, since they are generated using key constraints described in Section~\ref{subsec:resynth}.

\begin{figure*}[t]
	\centering
	\parbox{9.0cm}{\centerline{\includegraphics[width=9.0cm]{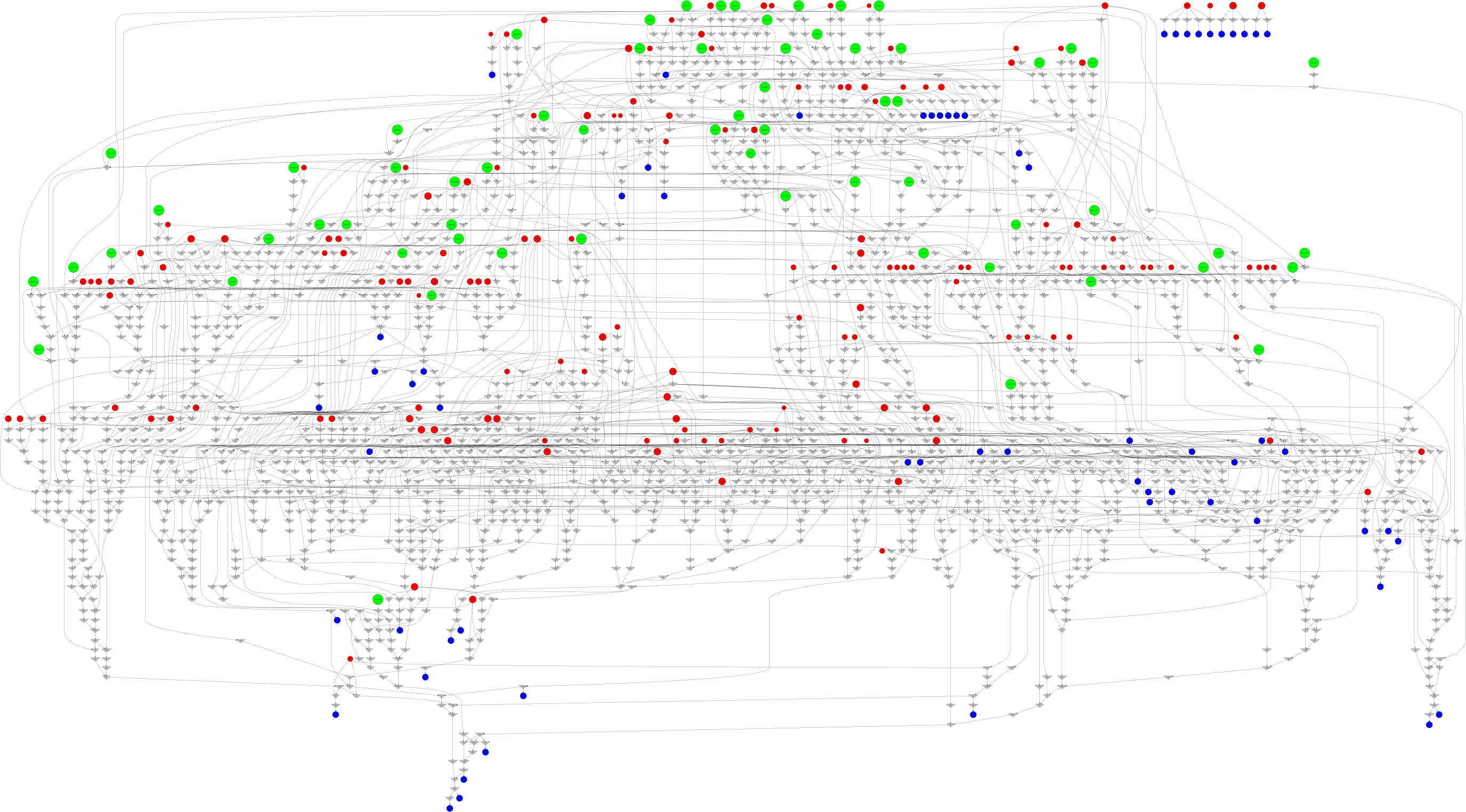}}}\
	\parbox{9.0cm}{\centerline{\includegraphics[width=9.0cm]{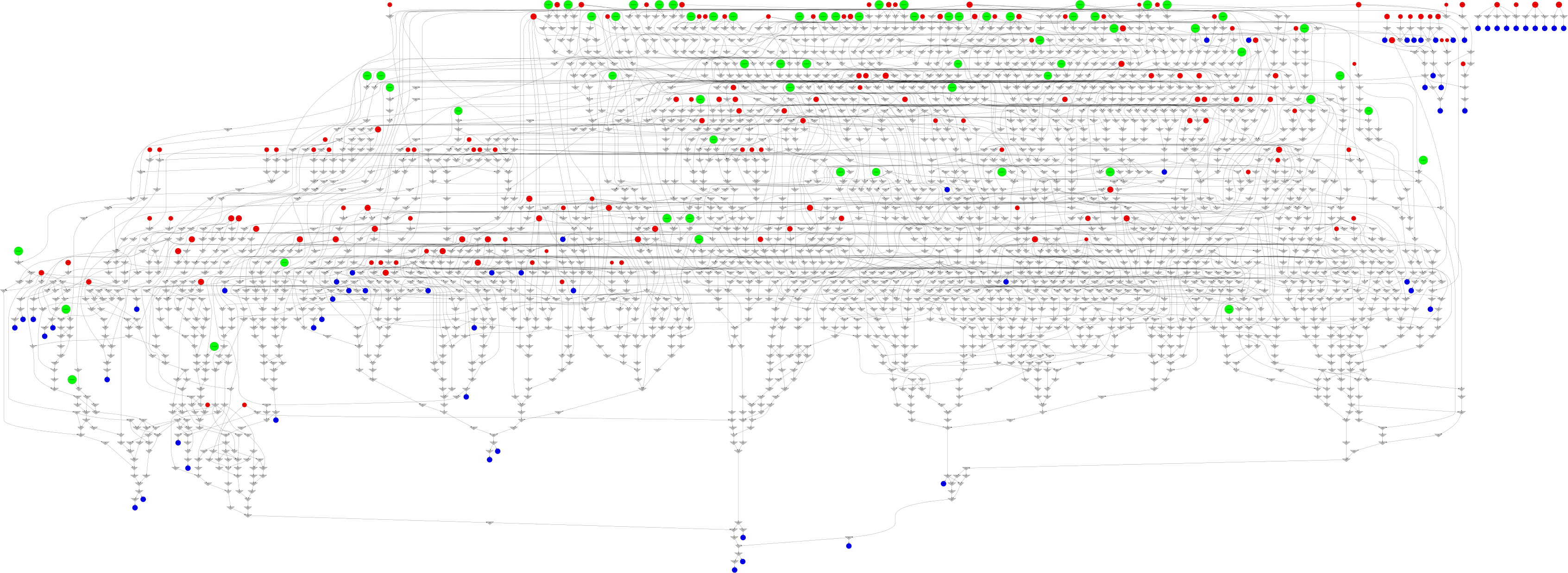}}}\
	\vspace*{1mm}
	\parbox{9.0cm}{\centerline{\footnotesize (a)}}\
	\parbox{9.0cm}{\centerline{\footnotesize (b)}}\
	\vspace*{-8mm}
	\caption{Graphs of resynthesized netlists generated using a difference in the delay constraint $dc$: (a) $dc$ is 990\,ps; (b) $dc$ is 496\,ps.}  
	\label{fig:structures}
        \vspace*{-4mm}
\end{figure*}

In order to illustrate the differences in the structure of generated netlists, the \textit{c2670} circuit locked by SKG-Lock is considered. Fig.~\ref{fig:structures} presents the graphs of two netlists resynthesized using the same synthesis parameters, except for the delay constraint. In this figure, red, green, and blue circles denote the inputs, key bits, and outputs, respectively; the gray triangles represent the gates. Observe that a small change in the delay constraint can lead to a structurally different netlist, where the difference between the number of gates and logic levels is 599 and 12, respectively.

\subsubsection{Attacks on the Locked ISCAS'85 Circuits}

Table~\ref{tab:attack-iscas85} presents the results of the SCOPE attack on the original locked netlists and those of OL resynthesis-based attack on the unique locked netlists generated in the resynthesis process. In this table, \textit{cdk} and \textit{dk} stand respectively for the number of correctly deciphered key bits and the total number of deciphered key bits and \textit{time} is the total time required for the attack. The attacks were also run on the same server used to resynthesize the locked netlists. 

Observe from Table~\ref{tab:attack-iscas85} that the SCOPE attack is not entirely successful on any of the original locked netlists. However, the use of resynthesized netlists enables us to decipher the values of a large number of key bits, and even the whole key, e.g., for the \textit{c2670} and \textit{c3540} circuits locked by SKG-Lock. Note that the SCOPE attack can decipher almost all of the key bits using the resynthesized netlists locked by the SKG-Lock technique. While the results on the netlists locked by SKG-Lock are all correct, the ones on the netlists locked by Anti-SAT, CASLock, and SFLL are slightly better than a random guess. The run time of the SCOPE attack and our resynthesis-based attack depends mainly on the number of gates and keys in the locked design. 

To find the SAT resiliency of resynthesized locked circuits, the SAT-based attack of~\cite{subramanyan2015} was run on 541 netlists of the \textit{c3540} circuit locked by SKG-Lock with a time limit of 2 days. This circuit was chosen since it has the smallest number of key bits with the smallest number of gates. Note that the SAT-based attack was not able to find the secret key of any resynthesized locked netlists. This experiment indicates that the resynthesis changes only the structure of the circuit as shown in Fig.~\ref{fig:structures}, but maintains its SAT resiliency. 

\subsubsection{Redundant Synthesis Runs}
\label{subsubsec:analysis}

Observe from Tables~\ref{tab:resynth-iscas85} and~\ref{tab:attack-iscas85} that the total run-time of the proposed attack is dominated by the resynthesis process. However, it is possible to reduce the time required to resynthesize the locked netlist by removing redundant synthesis runs without sacrificing any unique netlists. For example, it is observed that the \textit{high} value of the \textit{syn\_gen} parameter given in Section~\ref{subsec:resynth} can be removed from the parameter list, since all possible synthesis scripts including this parameter generate the same circuit when this parameter is \textit{low} or \textit{medium}. Thus, the number of generated circuits, i.e., 1080, reduces to 720.

\begin{table*}[t]
	\centering
	\footnotesize
	\caption{Results of Attacks on the locked ISCAS'85 Circuits.}
	\begin{tabular}{|@{\hskip3pt}l@{\hskip3pt}|c@{\hskip3pt}c@{\hskip3pt}|c@{\hskip3pt}c@{\hskip3pt}|c@{\hskip3pt}c@{\hskip3pt}|c@{\hskip3pt}c@{\hskip3pt}|c@{\hskip3pt}c@{\hskip3pt}|c@{\hskip3pt}c@{\hskip3pt}|c@{\hskip3pt}c@{\hskip3pt}|c@{\hskip3pt}c@{\hskip3pt}|}
		\hline
		\multirow{3}{*} {Circuit}  & \multicolumn{4}{c|}{Anti-SAT} & \multicolumn{4}{c|}{CASLock} & \multicolumn{4}{c|}{SFLL} & \multicolumn{4}{c|}{SKG-Lock}\\
		\cline{2-17}
		& \multicolumn{2}{c|}{SCOPE} & \multicolumn{2}{@{\hskip3pt}c|}{Resynthesis} & \multicolumn{2}{c|}{SCOPE} & \multicolumn{2}{c|}{Resynthesis} & \multicolumn{2}{c|}{SCOPE} & \multicolumn{2}{c|}{Resynthesis} & \multicolumn{2}{c|}{SCOPE} & \multicolumn{2}{c|}{Resynthesis}\\
		\cline{2-17}
		& cdk/dk & time & cdk/dk & time & cdk/dk & time & cdk/dk & time & cdk/dk & time & cdk/dk & time & cdk/dk & time & cdk/dk & time  \\
		\hline \hline
		c2670     & 0/0 & 4s & 37/64 & 34m18s & 0/0 & 4s & 35/64 & 33m47s & 0/0 & 4s  & 34/64 & 37m32s & 32/32 & 4s  & 64/64 & 44m37s \\
		c3540     & 0/0 & 3s & 17/32 & 21m27s & 0/0 & 3s & 17/32 & 18m12s & 0/0 & 2s  & 19/32 & 21m29s    & 17/17 & 2s  & 32/32 & 24m30s   \\
		c5315     & 0/0 & 5s & 38/64 & 42m34s & 0/0 & 5s & 30/64 & 43m54s & 0/0 & 5s  & 33/64 & 46m23s  & 32/32 & 5s & 62/62 & 52m06s \\
		c6288     & 0/0  & 3s & 18/32    & 29m08s     & 0/0  & 3s & 16/32    & 27m18s     & 0/0 & 3s  & 16/31 & 33m19s    & 16/16 & 3s  & 31/31 & 34m24s    \\
		c7552     & 0/0 & 6s & 38/64 & 45m31s & 0/0 & 6s & 47/64 & 49m13s & 0/0 & 6s & 38/63 & 52m26s & 32/32 & 6s & 61/61 & 56m45s  \\
		\hline
	\end{tabular}
	\label{tab:attack-iscas85}
        \vspace*{-4mm}
\end{table*}

\subsubsection{Convergence on the Number of Deciphered Keys}
\label{subsubsec:key}

\begin{figure}[t]
    \centering
    \includegraphics[width=8.5cm]{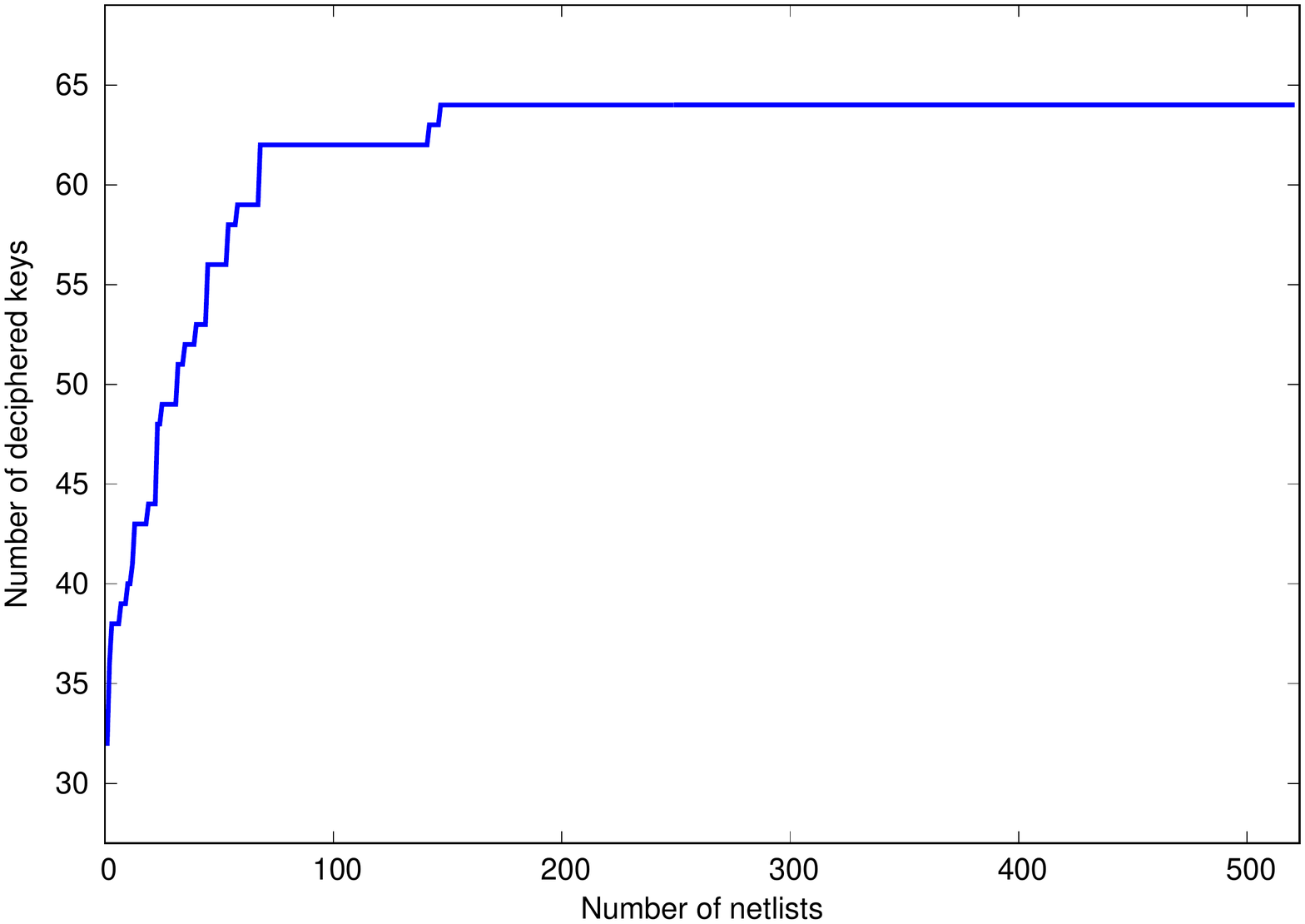}
     \vspace*{-8mm}
    \caption{Convergence on the number of deciphered keys over the number of resynthesized netlists in the SCOPE attack.}
    \label{fig:converge}
    \vspace{-4mm}
\end{figure}

It is also observed that the number of key bits deciphered by the SCOPE attack on all unique resynthesized netlists can actually be obtained using a small number of netlists. Fig.~\ref{fig:converge} presents the number of deciphered key bits along the unique resynthesized netlists of the \textit{c2670} circuit locked by \mbox{SKG-Lock}. Observe from this figure that although a large number of unique netlists increases the quality of the SCOPE attack, actually a small number of unique netlists, 147 in this case, is sufficient to achieve the same result as when all 521 unique netlists are considered. We note that a similar situation was also observed on circuits locked by other techniques. 

\subsubsection{Promising Resynthesized Netlists}
\label{subsubsec:slack}

Moreover, it is observed that the SCOPE attack is more successful on specific resynthesized netlists. To find a set of synthesis parameters that enables the SCOPE attack to decipher more key values, we initially define two categories of netlists based on the slack time of the design, i.e., the difference between the required and arrived time in the critical path, as follows: i)~netlists with a slack value less than or equal to 0; ii)~netlists with a slack value greater than 0. The slack value of a design gives indeed a rough idea of the effort put in by the synthesis tool; for the netlists in the first category, the synthesis tool works extremely hard to meet the delay constraint, trying many logic transformations and optimization techniques. 

Then, the solutions of the SCOPE attack on all possible 1080 netlists are obtained and sorted based on the number of deciphered key bits in descending order. The top 10\% of these sorted netlists are categorized based on their slack values.

\begin{figure}[t]
    \centering
    \includegraphics[width=8.5cm]{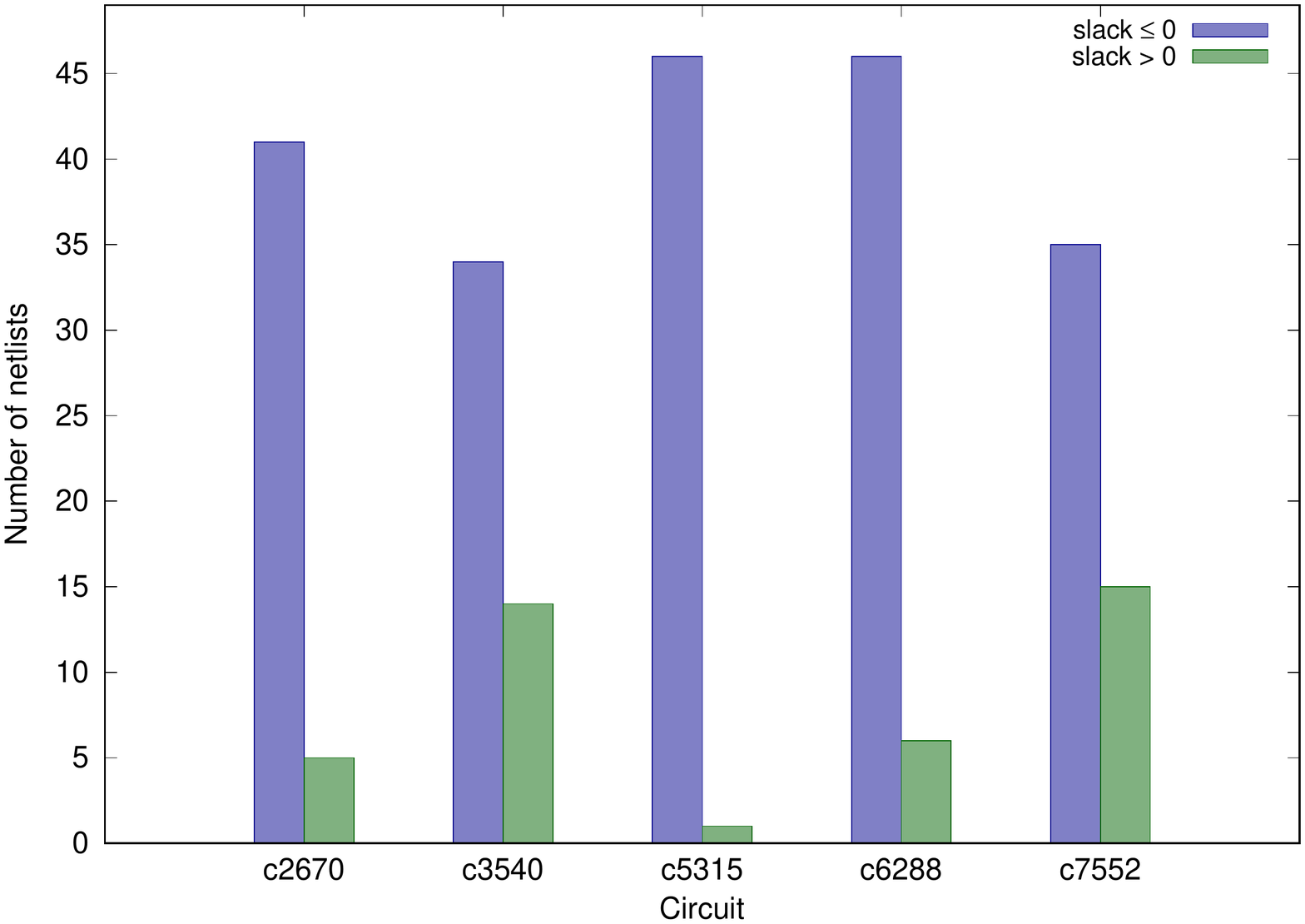}
     \vspace*{-8mm}
    \caption{Classification of resynthesized netlists based on their slack values on promising solutions of SCOPE attack.}
    \label{fig:category}
    \vspace{-4mm}
\end{figure}

Fig.~\ref{fig:category} presents the results of this experiment on the circuits locked by SKG-Lock. Observe that the netlists that enable the SCOPE attack to decipher more key values generally have a slack value less than or equal to 0. Thus, to generate such circuits, one can easily add strict delay constraints or key constraints as described in Section~\ref{subsec:resynth}. We note that a similar result was also observed on resynthesized netlists locked by other techniques.

\subsubsection{Structural Analysis}
\label{subsubsec:improvement}

In order to improve the performance of the resynthesis process, the logic cone, which is the locking technique is applied on, can be extracted and resynthesized. Note that the output of this logic cone is a single primary output, while its inputs are primary inputs, but not necessarily all the primary inputs of the locked design.  Thus, the run-time of the resynthesis process can be decreased, since the logic cone has a small number of inputs, outputs, and gates when compared to the whole locked circuit. 

%The tool generates a list of all outputs achieved from all the key inputs. Based on these outputs, all other outputs are disconnected using the command \textit{edit\_netlist}. Although the outputs are disconnected, generating a new netlist would keep these outputs floating. Therefore, another command is used, the \textit{delete\_unloaded\_undriven}. All these outputs are removed from the locked netlist, being generated a small netlist with only the inputs related to the outputs based on the key inputs.

Table~\ref{tab:optimization} presents details on the resynthesis process on entire locked circuits and logic cones when the circuits locked by SFLL are used. Observe that the resynthesis process on a logic cone generates less number of unique designs and takes significantly less time without a significant loss on the solution quality when compared to the resynthesis process on the entire circuit. We note that similar results were also observed on circuits locked by other techniques.

\begin{table}[t]
  \centering
  \footnotesize
  \caption{Results of the resynthesis process on entire circuit and logic cone. }
  \vspace{-2mm}
  \begin{tabular}{|@{\hskip3pt}l@{\hskip3pt}|c@{\hskip3pt}|@{\hskip3pt}c@{\hskip3pt}|@{\hskip3pt}c|c@{\hskip3pt}|@{\hskip3pt}c@{\hskip3pt}|@{\hskip3pt}c|}
    \hline
    \multirow{2}{*} {Circuit} & \multicolumn{3}{c|}{Entire Circuit} & \multicolumn{3}{c|}{Logic Cone} \\
    \cline{2-7}
    & \#unique & time & cdk/dk & \#unique & time & cdk/dk\\
    \hline \hline
    c2670 & 468 & 13h13m23s  & 34/64 & 319 & 07h46m26s & 34/64 \\ 
    c3540 & 484 & 1d47m51s   & 19/32 & 320 & 06h29m35s & 16/32 \\
    c5315 & 477 & 21h57m14s  & 33/64 & 313 & 07h06m16s & 32/64 \\
    c6288 & 523 & 2d22h15m7s & 16/31 & 302 & 06h20m57s & 19/32 \\
    c7552 & 504 & 22h40m29s  & 38/63 & 279 & 06h57m14s & 38/63 \\
    \hline
  \end{tabular}
  \label{tab:optimization}
\end{table}

\subsection{Results on the CSAW'19 Circuits}
\label{subsec:csaw19}

As the second experiment set, we used the state-of-the-art locked circuits from the CSAW'19 contest~\cite{tan20}. Details of these circuits are given in Table~\ref{tab:csaw19}. Note that two logic locking techniques -- RLL~\cite{roy2008} and SFLL-rem~\cite{sengupta20} -- are applied together to lock a circuit.

\begin{table}[t]
  \centering
  \footnotesize
  \caption{Details of the Locked CSAW'19 Circuits.}
  \vspace{-2mm}
  \begin{tabular}{|l|ccc|ccc|}
    \hline
    \multirow{2}{*} {Circuit} & \multicolumn{3}{c|}{Details} & \multicolumn{3}{c|}{Number of keys} \\
    \cline{2-7}
                              & \#in & \#out & \#gates & RLL & SFLL-rem & Total \\
    \hline \hline
    small  & 522  & 512  & 15995 & 40  & 40  & 80 \\
    medium & 767  & 757  & 24008 & 60  & 60  & 120 \\
    large  & 1452 & 1445 & 36584 & 80  & 80  & 160 \\
    bonus  & 892  & 1746 & 23004 & 128 & 128 & 256 \\
    \hline
  \end{tabular}
  \label{tab:csaw19}
\end{table}

In the following two subsections, we present the results of the resynthesis process and the resynthesis-based attack. 

\begin{table}[t]
  \centering
  \footnotesize
  \caption{Results of Resynthesized Locked CSAW'19 Circuits.}
  \vspace{-2mm}
  \begin{tabular}{|l|c|ccc|c|}
    \hline
    Circuit & unique & area & delay & power & time \\
    \hline \hline
    small  & 557 & 18935 & 1631 & 23571 & 5d3h22m28s \\ 
    medium & 569 & 26080 & 1745 & 31284 & 6d12h24m16s \\
    large  & 567 & 31348 & 1798 & 24610 & 5d21h42m10s  \\
    bonus  & 560 & 20643 & 1758 & 19090 & 4d14h44m29s \\
    \hline
  \end{tabular}
  \label{tab:resynth-csaw19}
\end{table}

\begin{table}[t]
	\centering
	\footnotesize
	\caption{Results of Attacks on the Locked CSAW'19 Circuits.}
        \vspace{-2mm}
	\begin{tabular}{|@{\hskip1pt}l@{\hskip1pt}|c@{\hskip1pt}c@{\hskip1pt}|c@{\hskip1pt}c@{\hskip1pt}|c@{\hskip2pt}c@{\hskip2pt}|}
		\hline 
		\multirow{2}{*} {Circuit-Netlist} & \multicolumn{2}{c|}{SCOPE} & \multicolumn{2}{c|}{Query} & \multicolumn{2}{c|}{Resynthesis} \\
		\cline{2-7}
		& dk & time & prv & time & dk & time\\
		\hline \hline
		small - OLN   & 19  & 20s       & 39  & 1m21s       & 40  & 1m41s       \\
 		small - URNs  & 77  & 4h10m42s  & 40  & 1d10h4m37s  & 79  & 1d14h15m19s \\
		\hline      
		medium - OLN  & 32  & 41s       & 58  & 6m37s       & 59  & 7m18s       \\
		medium - URNs & 117 & 8h33m56s  & 58  & 3d19h12m13s & 120 & 4d3h46m9s   \\
		\hline
		large - OLN   & 30  & 1m7s      & 79  & 6m19s       & 79  & 7m26s       \\
		large - URNs  & 152 & 12h56m15s & 80  & 3d2h52m11s  & 159 & 3d15h48m26s \\
		\hline
		bonus - OLN   & 64  & 1m46s     & 118 & 3m2s        & 120 & 4m48s       \\
		bonus - URNs  & 233 & 16h7m17s  & 125 & 1d20h29m22s & 252 & 2d12h36m39s \\
		\hline
	\end{tabular}
	\label{tab:attack-csaw19}
\end{table}

\begin{table*}[t]
	\centering
	\footnotesize
	\caption{Results of attacks on the locked CSAW'19 Circuits.}
        \vspace{-2mm}
	\begin{tabular}{|l|c|c|c|c|c|c|c|c|c|}
		\hline
		\multirow{3}{*}{Approach} & \multirow{3}{*}{Attack Scenario}  & \multicolumn{8}{c|}{Circuit} \\
		\cline{3-10}
		& & \multicolumn{2}{c|}{small (40+40)} & \multicolumn{2}{c|}{medium (60+60)} & \multicolumn{2}{c|}{large (80+80)} & \multicolumn{2}{c|}{bonus (128+128)} \\
		\cline{3-10}
		& &  {RLL} & {SFLL-rem} &  {RLL} & {SFLL-rem} &  {RLL} & {SFLL-rem} &  {RLL} & {SFLL-rem} \\
		\hline \hline 
		{Key sensitization~\cite{rajendran12}}       & {OG}                 & {40/40} & {---}   & {60/60} & {---}   & {80/80} & {---}   & {---}     & {---} \\
		{Hamming distance-based attack~\cite{tan20}} & {OG}                 & {30/30} & {---}   & {50/50} & {---}   & {72/72} & {---}   & {---}     & {---} \\
		{Automated analysis + SAT~\cite{tan20}}      & {OG}                 & {11/18} & {---}   & {31/50} & {---}   & {10/34} & {---}   & {---}     & {---} \\
		{Sub-circuit SAT~\cite{tan20}}               & {OG}                 & {17/17} & {---}   & {29/29} & {---}   & {---}   & {---}   & {---}     & {---} \\
		{Redundancy-based~\cite{li19}}               & {OL}                 & {28/28} & {4/12}  & {35/35} & {23/28} & {45/45} & {0/51}  & {66/66}   & {8/27} \\
		{Bit-flipping attack~\cite{shen18}}          & {OG}                 & {40/40} & {---}   & {60/60} & {---}   & {80/80} & {---}   & {---}     & {---} \\
		{Topology guided attack~\cite{zhang19}}      & {OL}                 & {15/32} & {---}   & {19/50} & {---}   & {36/73} & {---}   & {75/108}  & {---} \\
		\textbf{Resynthesis-based attack}   & {OG}     & {40/40} & {20/39} & {60/60} & {29/60} & {80/80} & {35/79} & {128/128} & {55/124} \\
		\hline
	\end{tabular}
	\label{tab:all-attacks-csaw19}
\end{table*}

\subsubsection{Resynthesis of the Locked CSAW'19 Circuits}

Table~\ref{tab:resynth-csaw19} presents the resynthesis results of locked circuits. Observe that the number of unique resynthesized netlists is larger than half of the total number of generated netlists, i.e., 540. As the hardware complexity of designs increases, the run-time of the resynthesis process increases. We note that diverse netlists in terms of complexity are obtained, e.g., the standard deviation on area, delay, and power dissipation values of all the locked netlists of the \textit{small} circuit is computed as 8526, 1029, and 20074, respectively.

\subsubsection{Attacks on the Locked CSAW'19 Circuits}

Table~\ref{tab:attack-csaw19} presents results of the attacks obtained, after they are applied to the original locked netlist, denoted as \textit{OLN}, and all unique resynthesized netlists, denoted as \textit{URNs}. In this table, \textit{prv} stands for the number of proven values of key bits. Note that since the secret key is \textbf{not publicly available}, the \textit{cdk} values are omitted for the SCOPE and resynthesis-based attacks.

Observe from Table~\ref{tab:attack-csaw19} that the original SCOPE attack could only decipher a small number of key bits, all of which belongs to RLL, and the query attack can prove the values of a large number of key bits, all of which again belong to RLL, on the original locked circuits. Thus, the resynthesis-based attack could only decipher the RLL key bits on the original locked circuits. However, the use of resynthesized circuits makes the SCOPE attack decipher more key bits that also belong to SFLL-rem and makes the query attack prove the values of more key bits that belong to RLL. Thus, the resynthesis-based attack could decipher almost all the values of the secret key, proving almost all the values of the key bits of RLL. Note that all the unknown key bits belong to SFLL-rem. Observe that the run-time of attacks increases, as the number of gates and key bits increases.

After the values of key bits of the CSAW'19 circuits were determined, they were sent to the contest organizers for evaluation. Table~\ref{tab:all-attacks-csaw19} presents the results of the resynthesis-based attack along with those of other techniques which participated in the contest. 

Observe from Table~\ref{tab:all-attacks-csaw19} that our proposed attack can determine all the key bits of RLL correctly, even though there are unproven key bits in the \textit{medium} and \textit{bonus} circuits as shown in Table~\ref{tab:attack-csaw19}. This observation implies that the guesses of the SCOPE attack on those key bits are actually correct. Moreover, the proposed technique can decipher the key bits of SFLL-rem with a number of deciphered key bits greater than any other OL technique with a high accuracy. 

	\section{Conclusions}
\label{sec:conclusions}

This work has shown that EDA tools can be used to generate variants of locked circuits that may be vulnerable to existing logic locking attacks and such circuits can be generated using a specific set of synthesis parameters. It was shown that the run-time of the proposed technique can be improved using a small number of resynthesized netlists without diminishing its solution quality. Experimental results clearly indicated that the use of many resynthesized circuits enables existing attacks to decipher values of a large number of key bits with high accuracy. Hence, the resynthesis of a locked circuit can be utilized as a pre-attack step for many existing attacks in order to improve their success rate. As future work, we plan to consider other synthesis parameters, such as fanout, capacitance limits, and wire loads, which enable synthesis tools to generate different circuits. Also, we aim to incorporate other commercial and open source EDA tools into the resynthesis process to generate different unique netlists. %Moreover, to make a strong guess on each key bit, we plan to use other OL attacks, such as redundancy-based~\cite{li19} and topology guided~\cite{zhang19}. 

	\section*{Acknowledgment}

The authors thank Nimisha Limaye for evaluating the keys found by the proposed technique on the CSAW'19 benchmarks.

	\bibliographystyle{IEEEtran}
	\bibliography{references}
\end{document}